\documentclass[article]{jss}
\usepackage{orcidlink, thumbpdf, lmodern}
\usepackage{amsmath, amssymb, bm, jlcode}
\usepackage{tabu, booktabs, float, tabularx}
\usepackage[noabbrev,capitalize]{cleveref}

\newcommand{\Ft}{\mathcal F_t}
\newcommand{\idx}[1]{\mathbb I \left\{#1\right\}}

\author{Rim Alhajal~\orcidlink{0009-0008-3010-5096}\\University of Aix-Marseille \And Oskar Laverny~\orcidlink{0000-0002-7508-999X}\\University of Aix-Marseille} 
\Plainauthor{Rim Alhajal, Oskar Laverny}
\title{\pkg{NetSurvival.jl}: A glimpse into relative survival analysis with \proglang{Julia}}
\Plaintitle{NetSurvival.jl: A glimpse into relative survival analysis with Julia}
\Shorttitle{\pkg{NetSurvival.jl}: relative survival analysis with \proglang{Julia}}
\Abstract{
  In many population-based medical studies, the specific cause of death is unidentified, unreliable or even unavailable.
  Relative survival analysis addresses this scenario, outside of standard (competing risks) survival analysis, to nevertheless estimate survival with respect to a specific cause.
  It separates the impact of the disease itself on mortality from other factors, such as age, sex, and general population trends.
  Different methods were created with the aim to construct consistent and efficient estimators for this purpose.
  The \proglang{R} package \pkg{relsurv} is the most commonly used today in application.
  With \proglang{Julia} continuously proving itself to be an efficient and powerful programming language, we felt the need to code a pure \proglang{Julia} take, thus \pkg{NetSurvival.jl}, of the standard routines and estimators in the field.
  The proposed implementation is clean, future-proof, well tested, and the package is correctly documented inside the rising \pkg{JuliaSurv} GitHub organization, ensuring trustability of the results.
  Through a comprehensive comparison in terms of performance and interface to \pkg{relsurv}, we highlight the benefits of the \proglang{Julia} developing environment.
}
\Plainkeywords{Net survival, cause of death, non-parametric estimation, performance, Julia} 
\Address{
  Rim Alhajal, Oskar Laverny\\
  INSERM, IRD, SESSTIM, Economic and Social Sciences of Health \& Medical Information Processing, ISSPAM\\
  University of Aix-Marseille\\
  Marseille, France\\
  E-mail: \email{rim.hajal@gmail.com}, \email{oskar.laverny@univ-amu.fr}\\
}

\begin{document}

\section{Introduction}
The field of survival analysis is targeted at the understanding, estimation, and modeling of time-to-event data, under censoring issues (\cite{andersen2012statistical}, \cite{fleming2013counting}, and \cite{Collett2023}).
Survival analysis has become crucial in medical research when understanding patients' survival given a specific scenario, such as a disease diagnosis.
Through statistical techniques, an estimation of the probability of the event (e.g., death) occurring over time is given, conditionally on the influence of various factors and covariates (\cite{Jenkins2005} and \cite{Survival2010}).
The analytical tools provided by the field help researchers analyze survival probabilities of a certain population or subgroup all the while identifying important variables that affect these probabilities.
For example, it allows the comparison between different treatments, and thus, establishing statistically which is more effective \citep{Treatments1994}.
Survival analysis is not restricted to human life times: it has a lot of industrial applications to assess the survival of components, or, e.g., the time to decay of particles.
By providing such insights, survival analysis grounds itself as a fundamental tool in all kinds of research, from applied statistics and biostatistics to chemistry, operational research and actuarial sciences.

There exists a comprehensive library of computational implementations of standard survival analysis tools, continuously growing and expanding.
This abundance is necessary to support ongoing biostatistical research, that leverages such computational routines daily.
For example, \proglang{R} packages like \pkg{survival} \citep{Therneau2015}, \pkg{flexsurv} \citep{jackson2016flexsurv}, \pkg{survminer} \citep{Kassambara2017}, \pkg{mexhaz} \citep{charvat2021mexhaz}, \pkg{xhaz} \citep{xhaz2022}, and \pkg{cmprsk} \citep{gray2014cmprsk} have contributed directly and heavily to the codebase.
They are leveraged regularly in applied research, e.g., most recently in a study pertaining to the excess hazard modelling applied to cancer epidemiology in \cite{elettiUnifyingFrameworkFlexible2022}, or in an investigation in machine learning models for predicting onset dementia using high-dimensional, heterogeneous clinical data from large cohort studies in \cite{spoonerComparisonMachineLearning2020}, among others. 

Data quality issues are prevalent in the field and can limit the direct application of standard survival analysis.
This is particularly the case in datasets coming from cancer registries where three different outcomes are envisioned for the patients: to be censored, to die from cancer, or to die from other causes \citep{jooste2013unbiased}.
When dealing with cancer data, the \emph{Excess mortality}, representing deaths by the studied cancer, is separated from the \emph{Population mortality}, representing all other causes of deaths.
While the reporting on censoring is usually exploitable, the reporting on the cause of death is, unfortunately, unreliable or straight-up unavailable \citep{PoharPerme2012}.
Out of this challenge, relative survival analysis was developed, thus, allowing researchers to isolate the impact of the specific event on survival from the population mortality (see \citep{gamel2001non} for original work on breast cancer). The main idea of the field is to leverage external data. In particular, census datasets can provide an estimation of the mortality rates of the general population. Then, through a few crucial hypotheses, estimation of the net survival can be performed.

The \proglang{R} package \pkg{relsurv} \citep{PermePavlik2018} has been the gold standard for non-parametric net survival analysis the past ten years.
It implements the most commonly used non-parametric estimators: Ederer 1 \citep{Ederer1961}, Ederer 2 \citep{Ederer1959}, and Pohar Perme \citep{PoharPerme2012}.
It also estimates the crude mortality rates, which give insight about the proportion of deaths related to the specific cause.
Furthermore, it estimates net sample size of subgroups as well as their expected average remaining lifetime \citep{andersen2017life}.
However, several issues arise from this implementation.
Due to the inherent slowness of interpreted languages such as \proglang{R}, and to the architecture of \pkg{relsurv} itself, the resulting code tends to have a relatively high execution time.
If this problem can be neglected when computing only a few survival curves, more computationally intensive use-cases are simply not possible on these grounds.
The code is difficult to read, even more to maintain, and barely tested.
The package does not include unit tests.
The internals are written in \proglang{C} and commented (sometimes wrongly) in several languages.
Nevertheless, the \pkg{relsurv} package remains the reference implementation in the field.
There also exists a \proglang{Stata} implementation (\cite{clerc2014net} from the same authors as \pkg{flexrsurv} \cite{clerc2021flexrsurv}, and \cite{lambert2009further}), sometimes used in research. For example, in \cite{allemani2015global}, they leverage cancer data from over 67 countries to establish a comprehensive global surveillance system for cancer survival, using the \proglang{Stata} implementation.
In this task, net survival serves as a measure of health system performance. 
However, since the \proglang{Stata} implementation is not open-source, and even paywalled, we are not able to discuss it further nor compare it to our \proglang{Julia} take.

While \proglang{R} remains a popular choice for statistical computing, \proglang{Julia} is gaining traction as a powerful alternative.
\proglang{Julia} is a high-performance language, built for scientific computing \citep{Bezanson2018}.
Its compiled nature allows its performance to match \proglang{C}, \proglang{C++} or \proglang{Fortran}, while providing an interface that is high level and featureful.
Its multiple dispatch paradigm boosts efficiency of the code, and allows interfaces to scale effortlessly for massive datasets and complex models.
The original manifest\footnote{\proglang{Julia}'s manifest: \url{https://julialang.org/blog/2012/02/why-we-created-julia/}} outlines the greedy philosophy underlying the language.
Remarkably, the language's success over the past 12 years reinforces the core principles described in that manifest.
\proglang{Julia}'s strong parametric typing ensures cleaner, more maintainable code — crucial for large scientific projects \citep{besanccon2019distributions}.
Its multiple dispatch paradigm \citep{bezanson2012julia} alongside its facilitated package management \citep{Pkg.jl} promotes code reuse within the \proglang{Julia} ecosystem \citep{bezanson2017julia}.
Though \proglang{R}'s vast package ecosystem and active community remain valuable, \proglang{Julia}'s focus on speed, performance, and scalability makes it a powerful option for researchers tackling demanding computational problems \citep{perkel2019julia}.
This adds to the motivation to contribute to extend the codebase available through \proglang{Julia}'s package registry.

The new \pkg{NetSurvival.jl} package provides non-parametric net survival estimation, alike \pkg{relsurv} but in pure \proglang{Julia}, with an ambitious interface and a remarkably faster run time.
It has a clean and compact codebase, an extensive collection of unit-tests, and an understanding documentation, which makes it future-proof.
To date, there is no other package in the \proglang{Julia} general registry that implements net survival estimators.
\pkg{NetSurvival.jl}'s development is part of an ongoing effort at the \code{JuliaSurv} GitHub organization to bring standard survival analysis tools into \proglang{Julia}.
A part of its internals, notably the fitting interface and the optimized rate tables fetching, overflowed from the original package into \pkg{SurvivalBase.jl} and \pkg{RateTables.jl} (resp.), created simultaneously by the same authors to provide unit dependencies for further developments.

This paper presents the underlying statistical and mathematical frameworks on which \pkg{NetSurvival.jl} is built on before focusing on the implementation itself and its potential.
Firstly, \cref{netsurv} provides a quick reminder on survival and relative survival analysis, and \cref{npe} describes the classic estimators from the later field.
Secondly, \cref{methodology} gives details on the technical implementation of both \pkg{RateTables.jl} and \pkg{NetSurvival.jl}, including, e.g., the classes hierarchies used, the rate tables internals and final API comparison with the \proglang{R/C} original implementation. It highlights in particular a few crucial \proglang{Julia} features.
Then, \cref{compar_r} provides a precise comparison with \proglang{R}, in terms of size of the code on one side and runtime on the other, highlighting the large performance boosts. 
Finally, \cref{example} contains a full real data example of use, and \cref{conclusion} concludes on the future of \pkg{NetSurvival.jl} inside \pkg{JuliaSurv}, the new \proglang{GitHub} organization that targets survival analysis in \proglang{Julia}.

\section{Relative survival analysis}\label{netsurv}

We start this section by the exposition of a formal set of notations, used to describe survival analysis and relative survival analysis problems. Our exposition is in line with standards in the field, in particular with \cite{fleming2013counting} and \cite{aalen2008survival}, which both contain a good introduction to the standard notations and theoretical developments in survival analysis.

For any positive random variable $X$, let us denote its survival function by $S_X(t) = \mathbb{P}(X > t)$ and its cumulative hazard function by $\Lambda_X(t) = -\ln S_X(t)$ for a given time $t$.
Furthermore, we denote by $\partial$ extended differentials of these functions, so that for a continuous random variable $X$, $\partial \Lambda_X$ denote its instantaneous hazard function and $-\partial S_X$ its density.

Consider two independent positives random variables $O$ and $C$, corresponding to the time before a certain event (e.g., death) and to the time until censorship respectively. Usually, the starting time corresponds to the time of diagnosis.
Then, denote $T = O \wedge C$ and $\Delta = \idx{T \le C}$ the random time to event and status of this event, respectively.
In standard survival analysis, we consider that only the $(T,\Delta)$ couple is observable, while the $(O,C)$ couple is not.
Let thus $(T_i,\Delta_i)_{i=1,...,n}$ be an observed i.i.d. $n$-sample of the $(T,\Delta)$ couple, indexed by $i \in 1,...,n$, and finally denote by $\left(\Omega,\mathcal A,\left\{\Ft, t\in \mathbb R_+\right\},\mathbb P \right)$ the associated filtered probability space, where the filtration corresponds to our observations: 
$$\Ft  = \sigma\left\{\left(T_i,\Delta_i\right): T_i \le t,\;\forall i \in 1,..,n\right\}.$$

On the one hand, the goal of standard survival analysis is to describe from these observations the distribution of the random variable $O$.
The distribution of the random variable $C$ is on the other hand considered a nuisance.
For that, the literature (e.g., \cite{aalen2008survival}) considers the stochastic processes $N_i(t) = \idx{T_i \le t, \Delta_i = 1}$ and $Y_i(t) = \idx{T_i \ge t}$, for all patients $i \in 1,...,n$, respectively dubbed the \emph{uncensored death processes} and the \emph{at-risk processes}. 
Remark that the previously defined filtration $\Ft$ can be rewritten as the natural filtration of $(N_i,Y_i)_{i \in 1,...,n}$:
$$\Ft = \sigma\left\{\left(N_i(s),Y_i(s)\right): 0 \le s \le t, i \in 1,...,n\right\}.$$

The standard non-parametric estimator for the distribution of $O$ is the Kaplan-Meier estimator \citep{kaplanNonparametricEstimationIncomplete1958}.
It first estimates the distribution of $O$ on the hazard scale, through the Nelson-Aalen estimator \citep{aalen1978nonparametric} whose expression at a given time $s$ is: 
$$\partial\hat{\Lambda}_O(s) = \frac{\sum\limits_{i=1}^n\partial N_i(s)}{\sum\limits_{i=1}^nY_i(s)},$$
and then uses a linearization of the exponential function around $0$ to obtain an estimator of the survival curve\footnote{There are other ways to obtain the Kaplan-Meier estimator, e.g., as a generalized maximum likelihood estimate.}: 
$$\hat S_O(t) =  \prod_{T_i \le t} \left(1 - \partial \hat{\Lambda}_{O}(T_i) \right) \approx \exp\left\{-\hat{\Lambda}_O(t)\right\}.$$

An estimation for the variance of the survival curve $\hat S_O(t)$ is provided using the Delta method \citep{cramer1946mathematical}, leveraging an estimation $\partial\hat\sigma_O^2$ of the variance $\sigma_O^2$ of $\hat{\Lambda}_O(s)$ given by the Greenwood formula \citep{greenwood1926natural}: 
$$\partial\hat\sigma_O^2(s) = \frac{\sum\limits_{i=1}^n\partial N_i(s)}{\left(\sum\limits_{i=1}^nY_i(s)\right) \left(\sum\limits_{i=1}^n(Y_i(s) - \partial N_i(s))\right)}$$

The construction of the estimators $\partial\hat\Lambda_0$ and $\partial\hat\sigma_0$ are justified by the counting processes framework, the Doob-Meyer decomposition of the uncensored death processes $N_i$, and the analysis of the resulting martingales. 
These estimators have a known and studied behavior, in both asymptotic and non-asymptotic settings.
The book \citep{fleming2013counting} contains a remarkable exposition of the theoretical analysis surrounding the few results we have just exposed.

In relative survival settings, we consider the time to death $O$ to be the minimum between two random times $E$ and $P$, $O = E \wedge P$.
Here, $E$ denotes the random time to death from the excess cause, and $P$ the random time to death from all other causes.
Due to censoring, all three of these variables, $E$, $P$, and $O$, are considered unobserved.
Due to data quality, as previously mentioned, the corresponding information being often unavailable or unreliable, the cause of death indicatrix $\idx{E \leq P}$ is also unobserved.
This missing indicatrix unfortunately holts the use of standard competitive risks methods.
When $E$ and $P$ are dependent random variables, these settings can lead to informative censoring \citep{turnbull1976empirical}.
If not all methods for estimating net survival takes this issue into account, the independence between $E$ and $P$ is a standard assumption in the field.

Another central assumption is the existence of reference mortality tables that provide the distribution of the general population mortality for each of the patients\footnote{Note that these mortality tables also include deaths due to the studied cancer that we consider negligible on a global scale. This is one of the limitations of the method: the studied cancer has to be rare for this approximation to work.}. 
Thus, we assume that the distribution of each $P_i$ is known.
However, due to demographic covariates (such as age at and date of diagnosis, the geographical region in which the patients live, etc.), we assume $P_1,...,P_n$ to be independent but not identically distributed.
Nevertheless, $E_1,...,E_n$ are fully i.i.d and independent of $P_1,...,P_n$.

The primary value of interest in net survival is the distribution of $E$, given by one of tits characteristic functions (density, survival curve, cumulative excess hazard or, equivalently, instantaneous excess hazard). 
The instantaneous excess hazard $\partial\Lambda_E$ is usually the main target of estimators in the field.
However, unlike Kaplan-Meier, most estimators for the excess hazard $\Lambda_E$ are not piecewise constant (which is logical since $\partial \Lambda_P$ is), and thus must be discretized at a very high rate for the linearization of the exponential to be justifiable.
The instantaneous excess hazard $\partial \Lambda_E$ can be estimated by several methods. We present the three most commonly used non-parametric ones in modern research in the following section.

\section{Non-parametric estimators}\label{npe}

The introduction of net survival analysis marked a significant step forward in understanding survival data related to specific diseases.
The field saw the development of numerous non-parametric estimators for the distribution of $E$ (\cite{Ederer1959}, \cite{Ederer1961}, \cite{Hakulinen1977}, etc…), some of which are implemented in the proposed package.
We present now a few of these estimators.

\subsection{Ederer 1}

Among the various methods for estimating survival, the Ederer 1 method \citep{Ederer1961} stands out for its simplicity and intuitive nature.
It directly compares the observed survival of the cohort to the survival from the general population, providing a clear interpretation of the relative impact of the given disease.
The estimator is given, for $s \ge 0$, by:
$$\partial\hat{\Lambda}_E(s) = \frac{\sum\limits_{i=1}^n \partial N_i(s)}{\sum\limits_{i=1}^nY_i(s)} - \frac{\sum\limits_{i=1}^nS_{P_i}(s)\partial\Lambda_{P_i}(s)}{\sum\limits_{i=1}^nS_{P_i}(s)},$$
and its associated variance is estimated by:
$$\partial\hat{\sigma}_E^2(s) = \frac{\sum\limits_{i=1}^n \partial N_i(s)}{\left(\sum\limits_{i=1}^nY_i(s)\right)^2}.$$

This method takes a unique approach in estimating the expected net survival.
Unlike some methods that consider changes in the group over time, Ederer 1 uses a group composition fixed at the beginning of the study.
Hence, patients that suffer early deaths are still considered part of the group when calculating population rates at later points, rendering it less computationally demanding \citep{esteve1990relative}.

In essence, Ederer 1 is a classic method that offers a consistent way to estimate the net survival.
While it has its limitations, it remains a valuable tool for researchers and a stepping stone to understanding more complex methods.
It is still used in studies, for example in the annual SEER Cancer Statistics Review (CSR) \citep{altekruse2010seer} which includes survival data from the United States and its respective analysis results from 1975 through 2007.

\subsection{Ederer 2}

In contrast to the previous method, the Ederer 2 method \citep{Ederer1959} takes a more dynamic approach.
It updates the number of individuals at risk at each time interval, reflecting the changing population structure due to events like deaths or withdrawals \citep{esteve1990relative}.
The estimator is given by:
$$\partial\hat{\Lambda}_E(s) = \frac{\sum\limits_{i=1}^n \partial N_i(s)}{\sum\limits_{i=1}^nY_i(s)} - \frac{\sum\limits_{i=1}^nY_i(s)\partial\Lambda_{P_i}(s)}{\sum\limits_{i=1}^nY_i(s)}.$$
The associated estimated variance is the same as for the Ederer 1 method.

\subsection{Pohar Perme}

The Pohar Perme estimator \citep{PoharPerme2012} is based on inverse probability of censoring weighting \citep{robins1993information}.
It accounts for informative censoring caused by the same demographic covariates influencing both non-disease-specific and disease-specific mortality times.
The estimator is given by:
\begin{equation}\label{eq:pohar_perme}
  \partial\hat{\Lambda}_E(s) =  \frac{\sum\limits_{i=1}^n\frac{\partial N_i(s)}{S_{P_i}(s)} - \sum\limits_{i=1}^n\frac{Y_i(s)}{S_{P_i}(s)}\partial\Lambda_{P_i}(s)}{\sum\limits_{i=1}^n\frac{Y_i(s)}{S_{P_i}(s)}},
\end{equation}
and its associated variance estimator is:
$$\partial\hat{\sigma}_E^2(s) = \frac{\sum\limits_{i=1}^n\frac{\partial N_i(s)}{S^2_{P_i}(s)}}{\left(\sum\limits_{i=1}^n\frac{Y_i(s)}{S_{p_i}(s)}\right)^2}.$$

Despite the fact that the inverse probability weighting is not based on probabilistic rationales, the Pohar Perme estimator from \cref{eq:pohar_perme} still has good properties.
According to \citep{PoharPerme2012} (see also \citep{seppa2016comparing}), it is biased but convergent and asymptotically unbiased.
The Pohar Perme estimator is now widely used in population-based studies.
In \cite{tron2023social} or \cite{grosclaude2017trends}, for example, the Pohar Perme estimator is applied to cancer registry data.
Simulation studies performed in \citep{danieli2012estimating} show a better performance (through net survival curve root mean square error) for the Pohar Perme estimator than for other known estimators of the net survival. 
A few more theoretical arguments (e.g., in \citep{PoharPerme2012}) lead us to prefer the use of this estimator over the others.
We nevertheless included both Ederer 1 and 2 estimators for academic comparison purposes.

\subsection{The log-rank type test}
To compare the survival distributions of two or more disjoint groups of individuals in standard survival analysis, a commonly used non-parametric test is the log-rank test, grounded on the same counting processes analysis as the Kaplan-Meier estimator \citep{fleming2013counting}.
For the comparison of net survival curves, a similar approach can be taken, following \cite{Graffeo2016}, to construct a log-rank-type test alongside the Pohar Perme estimator.
The null hypothesis $H_0$ for this test writes, for a given maximum time $T$:
$$\forall t \in [0,T], \; \; \Lambda_{E,g_1}(t) = \Lambda_{E,g_2}(t) = ... = \Lambda_{E,g_k}(t),$$
where $G = \{g_1,...,g_k\}$ is a partition of $1,...,n$ consisting of $k$ disjoint groups of patients that we wish to compare to each other.
The test is therefore designed to compare the hazard functions of the different groups, taking into account that they do not rely on exactly the same number of observations.
It allows for multiple groups, and to stratify the analysis on given covariates.

For all group $g \in G$, we denote the numerator and denominator of the Pohar Perme estimators restricted to individuals in the group, by: 
\begin{itemize}
  \item $\partial N_{E,g}(s) = \sum\limits_{i \in g} \frac{\partial N_i(s)}{S_{P_i}(s)} - \frac{Y_i(s)}{S_{P_i}(s)}\partial\Lambda_{P_i}(s)$
  \item $Y_{E,g}(s) = \sum\limits_{i \in g} \frac{Y_i(s)}{S_{P_i}(s)},$
\end{itemize}
and furthermore denote: 
\begin{itemize}
  \item $R_{g}(s) = Y_{E,g}(s) \left(\sum\limits_{g\in G} Y_{E,g}(s)\right)^{-1}$
  \item $Z_g(T) = N_{E,g}(s) - \int_{0}^T Y_{E,g}(s) \partial\hat{\Lambda}_E(s).$
\end{itemize}

Consider then the vector $\mathbf Z = \left(Z_{g}: \; g \in G \right)$.
It can be shown \citep{Graffeo2016} that it has asymptotically Gaussian distribution under the null hypothesis.
Thus, denoting $\hat{\bm\Sigma}$ an estimator of its variance, with entries
$$\hat\sigma_{g,h}(t) = \int_0^t \sum\limits_{\ell \in G} \left(\delta_{g,\ell} - R_g(s) \right)\left(\delta_{h,\ell} - R_h(s)\right) \left(\sum\limits_{i\in\ell} \frac{\partial N_i(s)}{S^2_{P_i}}\right),$$
the test statistic $U(T) = \mathbf Z(T)'\hat{\bm\Sigma}^{-1} \mathbf Z(T)$ becomes is asymptotically $\chi^2(k-1)$-distributed \citep{Graffeo2016} under $H_0$,  and can then be used to test the hypothesis.
There is also a stratified version of this test, which tests homogeneity in each group, while allowing heterogeneity between strata \cite{Graffeo2016}.

\subsection{Crude mortality}
The Pohar Perme estimator and its related log-rank tests focus on the hazard scale, a scale that brings a certain readability to the results due to the definition and properties of the instantaneous hazard itself.
Nevertheless, the scale of probabilities also brings interpretability, and several metrics on this scale were proposed in the literature.
This is in particular the case of the \emph{crude mortality} metrics \citep{kipourou2022direct}.
To define them, we decompose the overall mortality $(1 - S_O(t))$ into two components \citep{PermePavlik2018} by the law of total probabilities on the cause of death, obtaining: 
\begin{itemize}
  \item the \emph{crude probability of death from the disease}, given by $M_E(t) = \mathbb{P}(O \leq t, E \le P)$,
  \item the \emph{crude probability of death from other causes}, given by $M_P(t) = \mathbb{P}(O \leq t, P \le E)$.
\end{itemize}

Note that the crude probability of death from the disease should be interpreted with caution when comparing populations with disparate overall mortality rates, as it is clearly influenced by variations in population mortality.
In other terms, even if two cohorts show out the same excess mortality, there can still be a difference in the crude mortality depending on the population hazard given by the individuals in each cohort.

Both crude mortalities can be estimated from survival datasets where the cause of death is missing. Let us first derive expression for $M_E$ and $M_P$ that depend on other more standard functions:
\begin{align*}
M_E(t) &= \mathbb{P}\left(E \leq t, E \le P\right)\\
&= \int_0^\infty \mathbb{P}\left(e \leq t, e \le P\right) S_E(e-) \partial \Lambda_E(e)\\
&= \int_0^t S_P(e-) S_E(e-) \partial \Lambda_E(e)\\
&= \int_0^t S_O(e-) \partial \Lambda_E(e),
\end{align*}
and we also have, symmetrically, $M_P(t) = \int_0^t S_O(p-) \partial \Lambda_P(p)$.
Then, the standard method of estimation is a simple plug-in of estimators for the overall survival function and the two hazards. The resulting estimator is called the Cronin-Feuer estimator \citep{cronin2000cumulative}. It is given by: 
$$\hat{M}_E(t) = \int_0^t \hat{S}_O(u-)\partial\hat{\Lambda}_E(u),$$
where $\hat{S}_O$ is the Kaplan-Meier estimator and $\hat{\Lambda}_E$ is an estimated cumulative excess hazard (e.g., Pohar Perme).
An estimator of $\hat{M}_P$ can then simply be constructed by $\hat{M}_P = (1 - \hat{S}_O) - \hat{M}_E$ due to the previous discussion. 

\subsection{Expected net sample size}

One fundamental limitation of net survival is the high variability of estimators on long follow-up times \citep{manevski2023expected}.
As a reminder, net survival estimates the survival function of $E$ without taking into account the distribution of $P$.
Subsequently, estimating this for individuals unlikely to survive that long due to other causes (with respect to $P$) is impractical at best and nonsensical at worst.

The issue is that net survival is calculated through an average across all patients in the sample.
A stable estimate would require a sufficient number of people alive at the time point of interest.
Two things can be done: limit the follow-up time and/or restrict the analysis to subgroups that have higher chances of survival in the timeframe (e.g., younger patients or less-severe disease stages). 
This approach doesn't discard data, but rather focuses estimation on individuals for whom the information is meaningful.

To help assess the feasibility of long-term estimates, the concept of \emph{net sample size} can be used. 
The net sample size $n_E$ is defined as
$$n_E(t) = \sum\limits_{i=1}^n \idx{P_i > t}.$$
It indicates how many individuals are still at risk at a given time point, without accounting for excess morality.
Note that it does not account for censoring, and thus the true sample net sample size might be even lower.
However, the $P_i$ random variable are not observed and so this statistic cannot be observed as well. 
On the other hand, $P_i$'s distributions are known, and \citep{PermePavlik2018} proposed to look at the \emph{expected} net sample size $\mathbb E\left(n_E(t)\right)$, given by
$$\mathbb E\left(n_E(t)\right) = \sum\limits_{i=1}^nS_{P_i}(t).$$
Note that this function doesn't account for censoring.
As an additional reference, the expected mean remaining lifetime
$$\mathbb E\left(\frac{1}{n}\sum_{i=1}^n P_i\right)$$
is also provided within our implementation.
Developments related to the technical computation of this quantity are differed to \cref{sct:ratetables} which describes the link between the rate tables and the distributions of the random variables $P_1,..P_n$.

\section{Implementation}\label{methodology}

The implementation of the aforementioned estimators and functionalities started within the \pkg{NetSurvival.jl} package, but overflowed into several over packages inside the \code{JuliaSurv} GitHub organization. For example, \pkg{SurvivalBase.jl} hosts the fitting interface, and \pkg{RateTables.jl} holds the code and interface related to, not surprisingly, rate tables.
Both are useful building blocks for other parts of the ecosystem as well. The whole implementation is open-source and released under an MIT license. The following description is in line with its first published version.

\subsection[RateTables.jl]{\pkg{RateTables.jl}}\label{sct:ratetables}

In order to apply relative survival analysis, two different data types are needed: one cohort dataset and a population mortality table.
In \proglang{Julia}, the population mortality tables are \code{RateTable} objects from the \pkg{RateTables.jl} package.
This package includes a set of standard mortality tables from the literature. In particular, the rate table \code{slopop}, sourced from \pkg{relsurv} \citep{PermePavlik2018}, describes the population mortality in Slovenia by age, year of birth, and sex. We will use it later in our examples. 
It also includes a set of mortality tables taken from the Human Mortality Database\footnote{Human Mortality Database (HMD): \url{https://mortality.org/}}, in the \code{hmd\_rates} object. 

Loading the \pkg{RateTables.jl} package automatically imports a few \code{RateTable} objects such as \code{slopop} and \code{hmd\_rates}.
Showing these objects in the REPL gives information about the covariates names, and the \code{available\_covariates} helper function gives the corresponding values for a covariate.
You can access this information as follows:

\begin{lstlisting}[language=Julia]
julia> hmd_rates

RateTable(:country,:sex)

julia> available_covariates(hmd_rates, :sex)

(:female, :male, :total)
\end{lstlisting}

This particular rate table, \code{hmd_rates}, thus has two covariates related to the country and the sex of the person, and the sex covariate has three potential values: \code{:female}, \code{:male} and \code{:total}\footnote{The code \code{:something} in Julia produces a \code{Symbol} object, i.e., an \emph{interned string}, which for the purpose of our discussion can be thought as simply a string.}.
Country codes can be queried similarly.
Here we will simply use the \code{:svn} country code for Slovenia.
With this information, we can for example subset the \code{RateTable} object to obtain a \code{BasicRateTable} object.

\begin{lstlisting}[language=Julia]
julia> basicRT = hmd_rates[:svn,:male]

BasicRateTable:
ages, in years from 0 to 110 (in days from 0.0 to 40176.509999999995)
date, in years from 1983 to 2019 (in days from 724272.9029999999 to 737421.579) 
\end{lstlisting}

This allows us to acquire an instance of the \code{BasicRateTable} class, which serves as the foundation of the implementation.
These instances have very specific internal characteristics: they contain a matrix of daily hazard rates, indexed by age and date, both in yearly intervals\footnote{During the construction of these life tables, we meticulously addressed any irregularities to ensure they all conform to a common, imposed, yearly structure. Thus, the current implementation only provides yearly time steps on both axes. Nevertheless, a non-breaking update could add other possibilities, even non-regular ones, by creating another subclass of \code{AbstractRateTable}. This could be used to take into account more complicated data sources, at the cost of slower accessing routines.}. 

Formally, as implemented in the \code{BasicRateTable} class of the \pkg{RateTables.jl} package, a basic rate table simply encapsulate an \emph{offseted} matrix $\bm \lambda$ of daily hazard rates:
$$\bm \lambda = \left\{\lambda_{a,d} : a \in m_a,...,M_a; d \in m_d,...,M_d\right\},$$
where the first index is the age of the patient, and the second is the date. 

Ages and dates range between integers years $m_a < M_a$ and $m_d <M_d$ respectively. These four integers define the offsets of the rate matrix. 
Usual values for the extreme ages are $m_a = 0$ and $M_a = 103$ or $110$ or similar; a large enough age to cover human mortality.
Usual values for extreme dates are $M_d = 2020$ or $2024$, usually the year of extraction of the data, while $m_d$ can vary a bit more depending on the availability and quality of historical census data, e.g., from $1900$ to $1980$\footnote{Note that the standard implementation only allows for yearly intervals in the axes. Constructing life tables with another discretization while keeping the same outside interface would be easy, but we avoided it yet for the sake of simplicity of the proposed implementation.}.

A random life $P$, starting from age $a$ at a date $d$, follows the basic rate table $\bm\lambda$, if and only if it has instantaneous hazard function given, for $t \ge 0$, by: 
$$\lambda_P(t) = \lambda_{\textrm{clf}(a+t,m_a,M_a), \textrm{clf}(d+t,m_d,M_d)},$$ 
where the purposely built function $\textrm{clf}(x,m,M)$ simply clamps a number of \emph{days} $x$ between number of \emph{years} $m$ and $M$, and returns the number of integral years before the result.
It is formally defined as
$$\textrm{clf}(x,m,M) = \left\lfloor\min\left(\max\left(\frac{x}{365.241},m\right), M\right)\right\rfloor.$$

\begin{figure}[H]
  \centering
  \begin{minipage}{.5\textwidth}
    \centering
    \includegraphics{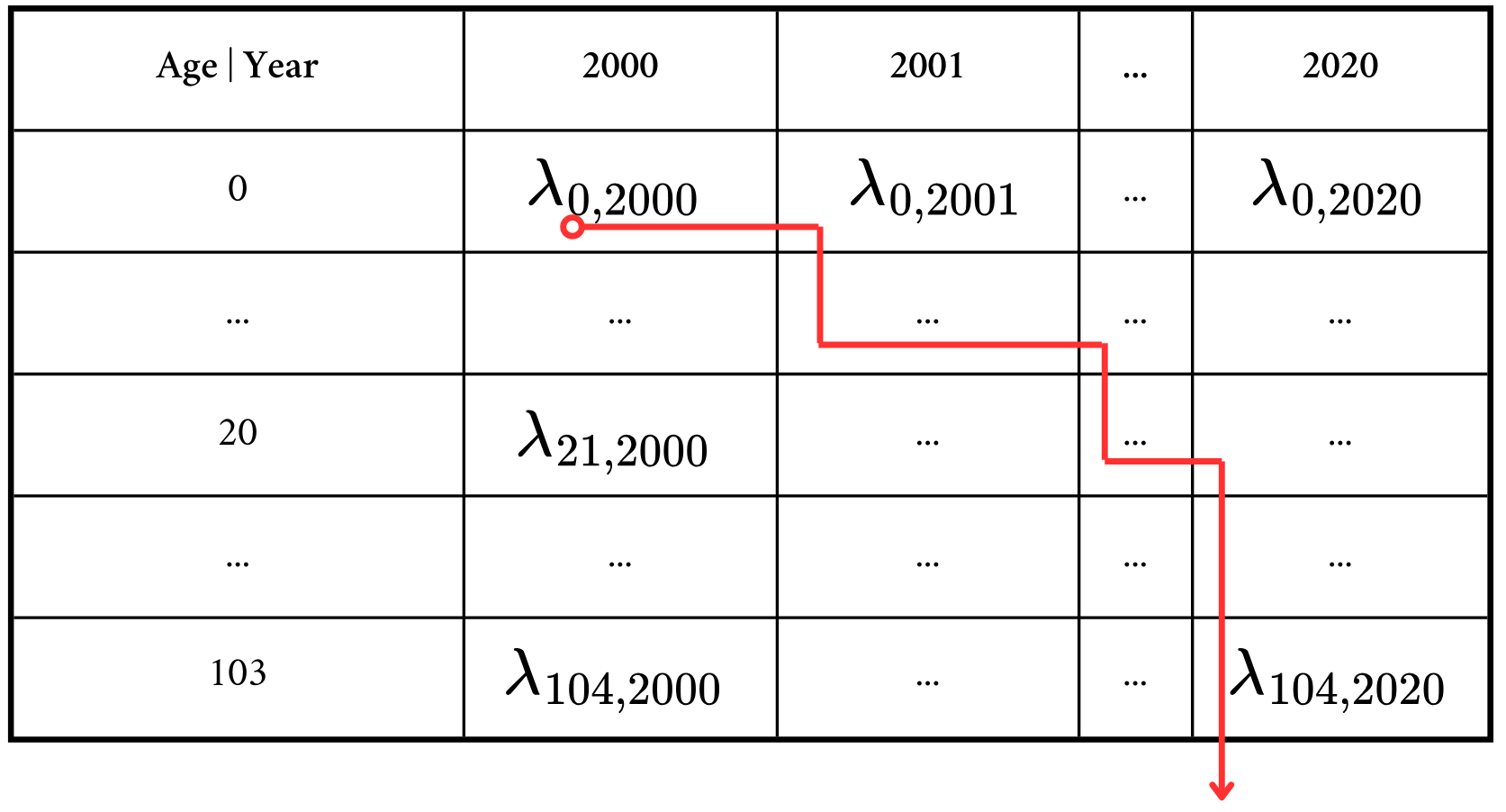}
  \end{minipage}%
  \begin{minipage}{.5\textwidth}
    \centering
    \includegraphics{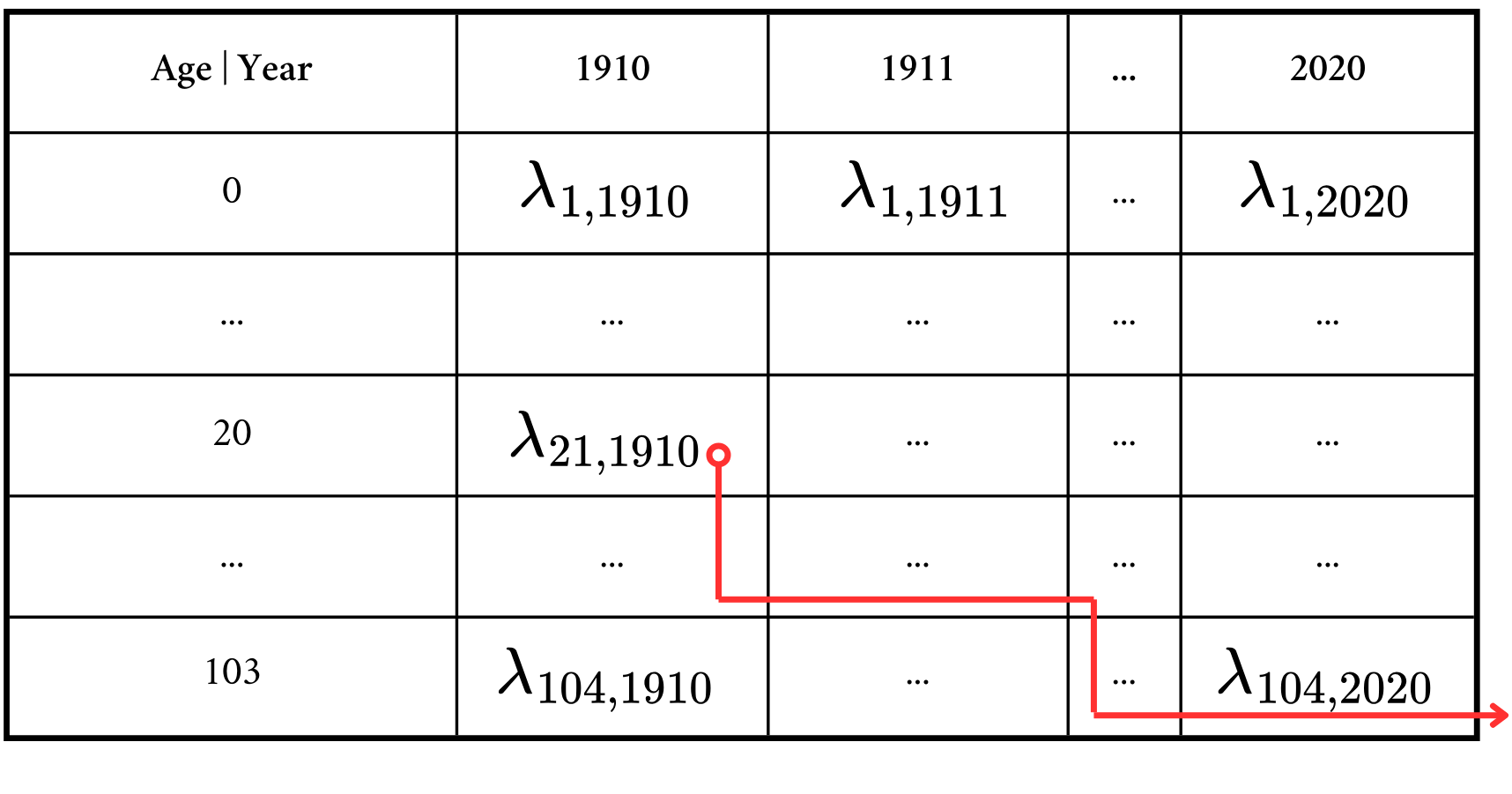}
  \end{minipage}
  \caption{Drawings of the path of two individuals in a basic rate table. On the left, an individual born in the year 2000 who reached the last year $M_d=2020$ of the rate table. On the left, an individual aged 20 in the year 1910 who reached the last age $M_a = 103$. The upper clamping mechanism will imply very specific behavior for these two individuals.}
  \label{fig:charts}
\end{figure}

This clamping behavior implements elegantly a very specific approximation scheme for out-of-bounds requests, as illustrated on \cref{fig:charts}.

In the example shown in \cref{fig:charts}, we consider two lives, on the same basic rate table.
For the person born in 2000, aged 20 in 2020, their path in the rate table is drawn on the left.
The clamping mechanism takes into account that this person is not dead at age $20$, and continues to follow the life table by going straight down.
We use the last column of the life table to represent its future mortality. 
On the other hand, the example on the right of \cref{fig:charts} shows a person exiting the life table from the bottom, hence at age 103.
The clamping mechanism considers here that time will continue to move on, but this person will always have terminal rates and therefore will follow the bottom line to the right until reaching the last cell.
The last cell's rate is applicable for eternity in both cases. In other terms, very old lives follow exponential distributions with this rate, which, since the rate should be quite big, is an okay approximation.

From the definition of the instantaneous hazard function, we can extrapolate to get the survival function, the density, random number generators, etc.
This extrapolation is implemented quite smoothly using a struct \code{Life<:Distributions.ContinuousUnivariateDistribution}, leveraging the random variable API from \pkg{Distributions.jl} directly.

On the other hand, since the rate table characterizes the distribution of $P$, the life expectation $\mathbb E(P)$ can be calculated exactly \citep{andersen2017life}.
For that, remark that there exists time points $t_0=0 < t_1 < ... < t_{n-1} < t_n = \infty$ forming a partition of $\mathbb{R_+}$ on which the instantaneous hazard function $\lambda_P(u)$ is piecewise constant, with values say $\lambda_j$'s:

\begin{align}
  \lambda_P(u) = \sum\limits_{j=1}^n \lambda_j \idx{u \in  \left]t_{j-1}, t_j\right]}.
  \label{eq:lambda}
\end{align}

The values of these times points depend on (and can be deduced from) initial age and date $(a,d)$ for the individual. Then, 
\begin{align*}
  \mathbb{E}(P) &= \int_0^\infty S_P (t) \partial t \\
  &= \sum\limits_{j=1}^n \int_{t_{j-1}}^{t_j}S_P (t)dt \\
  &= \sum\limits_{j=1}^n S_P (t_{j-1})\int_{t_{j-1}}^{t_j} \frac{S_P (t)}{S_P (t_{j-1})} dt.
\end{align*}

Each of these integrals can be simplified as follows:
\begin{align*}
  \int_{t_{j-1}}^{t_j} \frac{S_P (t)}{S_P (t_{j-1})} dt
  &= \int_{t_{j-1}}^{t_j} \frac{\exp\left(-\int_0^t \lambda_P(u)du \right)}{\exp\left(-\int_0^{t_{j-1}} \lambda_P(u)du \right)} dt\\
  &= \int_{t_{j-1}}^{t_j} \exp\left(-\int_{t_{j-1}}^{t_j}\lambda_P(u)du\right)dt \\
  &= \int_{t_{j-1}}^{t_j} \exp \left(-\left(t_j-t_{j-1}\right)\lambda_j\right)dt\\
  &= \frac{1-\exp \left(-\left(t_j-t_{j-1}\right) \lambda_j\right)}{\lambda_j}.
\end{align*}

Thus,
\begin{align}
  \mathbb{E}(P) = \sum\limits_{j=1}^n \frac{S_P(t_{j-1})}{\lambda_j}\left(1-\exp\left(-\left(t_j-t_{j-1}\right)\right)\right),
  \label{eq:expectation}
\end{align}
where $S_P(t_j) = \prod_{k=1}^{j} \exp\left\{-(t_k - t_{k-1})\lambda_k\right\}$ can be efficiently computed step by step alongside the sum. Since the number $n$ of cells that the person will go through is limited to about two times the remaining years to live, usually we have $n \ll 200$ and the computation of \cref{eq:expectation} is remarkably fast.

When life tables are provided yearly, in particular in the case of the Human Mortality Database, daily hazards rates are defined according to a piecewise constant hazard hypothesis from life tables yearly mortality probabilities. More precisely, if we denote $q_{a,d} = \mathbb P(T_{a,d} \le 1)$, where $T_{a,d}$ is the random lifetime, in years, of a person aged $a$ at date $d$, then the corresponding daily hazard rate $\lambda_{a,d}$ is computed through: 
$$\lambda_{a,d} = \frac{-\log \left(1 - q_{a,d}\right)}{365.241}.$$

We see on this bijection that we consider years that are $365.241$ days long. Note that other sub-yearly hypothesis could be used, you may see, e.g., \pkg{MortalityTables.jl} for a few other possibilities.

The primary functionality of these \code{RateTable} and \code{BasicRateTable} objects is querying mortality rates through \cref{eq:lambda}, which can be done using the \code{daily\_hazard} function.
It can be called in a variety of ways, but the arguments should follow a specific format.
The \code{age} and \code{date} parameters should be in days with the already mentioned conversion rate of $1$ year $=365.241$ days.
The format of other covariates may vary between rate tables, but it's essential to consider that their order is significant.

The following call for example queries a daily hazard rate for a Slovenian male, on his 20th birthday which happens to fall on the tenth of January 2010:

\begin{lstlisting}[language=Julia]
julia> daily_hazard(hmd_rates, 20 * 365.241, 2010 * 365.241 + 10, :svn, :male)
1.8021431794632215e-6
\end{lstlisting}

An equivalent syntax would be \code{daily_hazard(hmd_rates[:svn, :male], 20 * 365.241, 2010 * 365.241 + 10)}.
Retrieving these daily hazards is a highly sensitive operation that is carefully optimized considering how often it is used within critical loops (e.g., the Pohar Perme estimator).

\subsection[NetSurvival.jl]{\pkg{NetSurvival.jl}}

The \pkg{NetSurvival.jl} package on the other hand contains the main functionalities for relative survival.
It basically implements the non-parametric estimators described in \cref{npe}, leveraging deeply the API of \pkg{RateTables.jl}.

In \pkg{NetSurvival.jl}, the \code{NPNSE} parametric type\footnote{In \proglang{Julia}, classes are called \emph{types}, and are defined by the keyword \code{struct}. They are composite (object-oriented) data types that group together variables under a single name. The produced type, on which methods dispatch, can be parametrized. This is a powerful coding paradigm that tends to increase code reuse. Parametric typing is the possibility to create generic type families composed of several subtypes on which functions can dispatch and be specialized.
It is used extensively and in a calculated manner all across the package ecosystem, and in particular in \pkg{NetSurvival.jl}, to ensure minimal code redundancy.}, for \emph{Non-Parametric Net Survival Estimator}, allowed us to group the different non-parametric estimators under one type, and have them share a large part of their implementation (through parametric typing), from internal to public methods. For example, the mapping from $\partial \Lambda$ to $S$ (to obtain the survival curve alike Kaplan-Meier) was implemented only once as well as the allocation of workspace, the fitting interface, etc. By sharing a large portion of the code between methods through making certain choices at compile time, we increased efficiency and potential code reuse. This is almost invisible to the user thanks to the zero-cost abstraction that are the aliases.

In \cref{tab:functions_map} below, we can find the different functions that exist in both the R and Julia versions, comparing their interfaces to one another.

\begin{table}[H]
  \small
  \centering
  \begin{tabu}to \linewidth {>{\raggedright}l>{\raggedright}X}
    \toprule
    Description: & Command to compute a net survival curve with the Pohar Perme estimator\\
    \proglang{R} code: & \code{rs.surv(Surv(time, status) $\sim$ 1, colrec, slopop)}\\
    \proglang{Julia} code: & \code{fit(PoharPerme, @formula(Surv(time, status) $\sim$ 1), colrec, slopop)} \\
    \addlinespace\midrule\addlinespace
    Description:& Command to compute a net survival curve using another estimator (Ederer 1)\\
    \proglang{R} code: & \code{rs.surv(Surv(time, status) $\sim$ 1, colrec, slopop, method = "ederer1")}\\
    \proglang{Julia} code: & \code{fit(EdererI, @formula(Surv(time, status) $\sim$ 1), colrec, slopop)} \\
    \addlinespace\midrule\addlinespace
    Description:& Command to test net survival curve differences\\
    \proglang{R} code: & \code{rs.diff(Surv(time, status) $\sim$ sex, colrec, slopop)}\\
    \proglang{Julia} code: & \code{fit(GraffeoTest, @formula(Surv(time, status) $\sim$ sex), colrec, slopop)}\\
    \addlinespace\midrule\addlinespace
    Description:& Command to estimate the net expected sample size\\
    \proglang{R} code: & \code{nessie(Surv(time, status) $\sim$ sex, colrec, slopop)}\\
    \proglang{Julia} code: & \code{nessie(@formula(Surv(time, status) $\sim$ sex), colrec, slopop)}\\
    \addlinespace\midrule\addlinespace
    Description:& Command to compute the crude probability of death\\
    \proglang{R} code: & \code{cmp.rel(Surv(time, status) $\sim$ 1, colrec, slopop)}\\
    \proglang{Julia} code: & \code{fit(PoharPerme, @formula(Surv(time, status) $\sim$ 1), colrec, slopop) |> CrudeMortality}\\
    \bottomrule
  \end{tabu}
  \caption{Comparison of call-out interfaces for a few functionalities present in both \pkg{relsurv} and \pkg{NetSurvival.jl}, showcased using the \code{colrec} \& \code{slopop} datasets.}
  \label{tab:functions_map}
\end{table}

There isn't a big difference when comparing both interfaces allowing new users to get accustomed to the \proglang{Julia} version easily. Although subjective, it is fair to say the \pkg{NetSurvival.jl} interface is more transparent and straightforward, notably in the naming of the functions (e.g., \code{cmp.rel} compared to \code{CrudeMortality}). 

\section[Comparison with R]{Comparison with \proglang{R}}\label{compar_r}

Leveraging the \code{RateTable} objects API, \code{NetSurvival.jl} then implements the different estimators outlined in the \cref{npe}. One interesting difference between the \proglang{Julia} and \proglang{R} implementations is in their code length: the \proglang{Julia} version is more concise, as the \code{cloc}\footnote{The \code{cloc} software is a very small open-source utility to count the number of line of code in different languages in a repository. It can be found at \url{https://github.com/AlDanial/cloc}, and simply used by running the \code{cloc} command in a normal shell, positioned in the repository to be analyzed.} software output in \cref{tab:relsurv_lines} and \cref{tab:netsurvival_lines} shows.

\begin{table}[H]
  \small
    \centering
    \begin{tabu} to \linewidth {>{\raggedright}X>{\raggedright}X>{\raggedright}X>{\raggedright}X>{\raggedright}X}
    \toprule
    Language   & files     &     blank        &comment       &    code  \\
    \midrule
    R          &                     40      &      883     &      1793     &      6096\\
C                &               11      &      284     &       526    &       1222\\
C/C++ Header           &          1          &   11       &       6       &      25\\
\midrule
Total            &                52       &    1178     &      2325      &     7343\\
\bottomrule
    \end{tabu}
    \caption{Summary of the count of lines of code for different languages in \code{relsurv} obtained with the open-source \code{cloc} software.}
    \label{tab:relsurv_lines}
\end{table}
\begin{table}[H]
  \small
    \centering
    \begin{tabu} to \linewidth {>{\raggedright}X>{\raggedright}X>{\raggedright}X>{\raggedright}X>{\raggedright}X}
    \toprule
    Language   & files     &     blank        &comment       &    code  \\
    \midrule
    Julia & 16 & 151 & 143 & 547 \\
\bottomrule
    \end{tabu}
    \caption{Summary of the count of lines of code in \code{NetSurvival.jl} obtained with the open-source \code{cloc} software.}
    \label{tab:netsurvival_lines}
\end{table}

It is evident when comparing \cref{tab:relsurv_lines,tab:netsurvival_lines} that \pkg{NetSurvival.jl} has a big advantage in terms of number of files and lines of code ; for the same functionalities, it is around $15$ times smaller w.r.t lines of code. This is partially due to the expressiveness of \proglang{Julia} code, and the carefully planned object-oriented implementation. Additionally, it relies purely on one programming language, whereas the \proglang{R} implementation uses \proglang{C} as well. 

The coding experience is significantly easier when dealing with only one sole language, especially in the context of performant loops and interfaces. \proglang{Julia}'s strength here is in its standard tools\footnote{The existence of macros such as \texttt{@code\_warntype}, \texttt{@profview}, \texttt{@check\_allocs}, etc. facilitates the analysis of code and the optimization for performance}. Some tools allowed us to check the quality of our code (no type instabilities, entirely compiled, etc.). Others pinpointed the precise lines that delayed our code (e.g., the \code{daily_hazard} function, which led to us handling this particular feature with special care and implementing it as efficiently as possible). The functions for the net survival estimators are pretty optimal in that sense as \proglang{Julia} provided us with the necessary baggage leading to up to 60\% of their run time spent in computing the \code{exp} function. 

Another important difference between \pkg{NetSurvival.jl} and \pkg{relsurv} is bare speed.
In fact, for the computation of net survival curves, whatever the estimator used, runtime improvements are significant with respect to \pkg{relsurv}.
Computed on an i9-13900 processor, we present in \cref{timings} runtime multipliers comparing \proglang{Julia} with \proglang{R}, for this particular dataset on stratified and unstratified versions of several standard net survival routines.
\begin{table}[H]
  \small
  \begin{tabu} to \linewidth {>{\raggedright}X>{\raggedright}X>{\raggedright}X}
\toprule
$\partial \hat\Lambda_E$ method &  Unstratified & Stratified \\
      \midrule
      Pohar Perme  & 20.8431 &  20.1461 \\
      Ederer 1 & 7.216 &  4.1363 \\
      Ederer 2 & 29.2397 & 29.0399 \\
      \bottomrule
  \end{tabu}
  \caption{\label{timings} Runtime multipliers comparing \pkg{relsurv} to \pkg{NetSurvival.jl}, computed on a i9-13900 processor.
  The data used is the \code{colrec} dataset and the \code{slopop} mortality table, and potential stratification is done on \code{stage}.}
\end{table}

We see on \cref{timings} that the \proglang{Julia} versions are about $20$ to $30$ times faster, depending on the method used. This can quickly become significant if the amount of computations needed is large enough.

The same can be said for the log-rank type tests function. By comparing the speed of execution of these tests between the two implementations, the result is that, again, \pkg{NetSurvival.jl} is significantly faster, as seen in \cref{tbl:graf_rt}, on both stratified and unstratified circumstances. 
\begin{table}[H]
  \small
  \begin{tabu} to \linewidth {>{\raggedright}X>{\raggedright}X>{\raggedright}X}
\toprule
      &  Unstratified & Stratified \\
      \midrule
      Grafféo log-rank test & 13.1556 &  18.156 \\
      \bottomrule
  \end{tabu}
  \caption{Runtime multipliers comparing the Grafféo's log-rank test between \pkg{relsurv} and \pkg{NetSurvival.jl}, both in stratified and unstratified conditions.
  The data used is the \code{colrec} dataset and the \code{slopop} mortality table.}
  \label{tbl:graf_rt}
\end{table}

We found that the runtime improvements between the two implementation is in both cases (at least partially) coming from the \code{RateTable}'s \code{daily_hazard} fetching algorithm. Indeed, as implemented in the \pkg{survival} package, the query of population hazard rates uses a rather complex and slow \proglang{C} implementation, while the \proglang{Julia} implementation is much faster, while being also much shorter, easier to read and to maintain.

The \code{nessie} function is about 26 times faster on this example in \pkg{NetSurvival.jl} than its \proglang{R} equivalent.
This is again due to fast implementations of rate tables fetching in \pkg{RateTables.jl}, but most importantly to the fast \code{Life} function in \pkg{RateTables.jl}, as seen in \cref{sct:ratetables}.

Lastly, for the function outputting the crude mortality rates, the runtime difference is extremely large: our implementation is around $13000$ times faster as \cref{tbl:crude_mortality_speed} shows.
The comparison is a bit unfair though, because the \proglang{Julia} implementation reuses directly the already-computed net survival estimator instead of re-computing it.
However, in \pkg{relsurv}, only one method is proposed for the estimator or net survival used in Cronin-Feuer (the Ederer 1 method) whereas \pkg{NetSurvival.jl} gives a choice to the user and provides results that are in line with the estimator used. This also shows in the interface use for the functionality. 
\begin{table}[H]
  \small
  \begin{tabu} to \linewidth {>{\raggedright}X>{\raggedright}X>{\raggedright}X>{\raggedright}X}
\toprule
      &  PoharPerme & Ederer 1 & Ederer 2 \\
      \midrule
      Crude Mortality & 12022.7 &  13955.4 & 23938.8 \\
      \bottomrule
  \end{tabu}
  \caption{Runtime multipliers comparing the crude mortality runtimes between \code{relsurv} and \code{NetSurvival.jl}, computed on an i9-13900 processor.
  The data used is the \code{colrec} dataset and the \code{slopop} mortality table. Note that the \proglang{R} version only gives a result for the \code{EdererI} estimator, which timing we use to compare the other \proglang{Julia} versions too.}
  \label{tbl:crude_mortality_speed}
\end{table}

\section{Example}\label{example}
Let's apply \pkg{NetSurvival.jl}'s functionalities to a real example.
We use the \code{colrec} dataset (included in \pkg{NetSurvival.jl}) alongside the \code{slopop} mortality table (included in \pkg{RateTables.jl}) as reference. As the previous results, all graphs and results in this part are reproducible using the supplementary material published with this paper.

\begin{figure}[H]
  \centering
  \includegraphics[width=\textwidth]{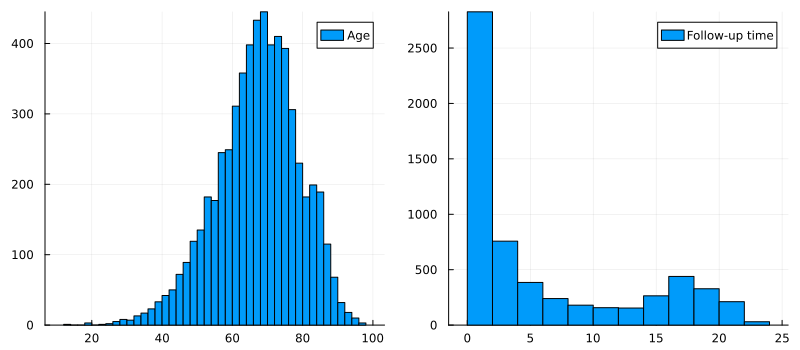}
  \caption{Age and follow-up time distribution in the \code{colrec} dataset.}
  \label{colrec_plot}
\end{figure}

The \code{colrec} dataset consists of $5971$ Slovenian patients diagnosed with colon or rectal cancer between $1994$ and $2000$.
There are $7$ different variables to describe said patients.
The three variables \code{age} (for the age at diagnosis in days), \code{year} (for the year of diagnosis in days), and \code{sex} (with options male or female) are used to query population data in the rate table.
The four other variables are \code{time} (follow-up time in days), \code{status} (uncensored indicatrix), \code{stage} (coding the cancer stages as 1, 2, 3, or 99 for unknown stage), and lastly the cancer \code{site} (with options rectum or colon).
To better describe the data at hand, \cref{colrec_plot} shows the distribution of ages and follow-up times in the dataset.

\Cref{colrec_plot} shows that a significant portion of the cancer patients in the study are aged between 60 and 80 years old.
Additionally, the follow-up time is less likely to surpass the 10-year mark.
Let's look closer into this phenomenon. To obtain a non-parametric estimate of the net survival from the \code{colrec} dataset and its associated mortality table \code{slopop}, we applied the Pohar Perme estimator using the following syntax:
\begin{lstlisting}[language=Julia]
using NetSurvival, RateTables, DataFrames
pp = fit(PoharPerme, @formula(Surv(time,status)~1), colrec, slopop)
\end{lstlisting}

You may simply replace \code{PoharPerme} by \code{EdererI} or \code{EdererII} depending on the desired method.
You can see in this call that the \code{Surv} function and \code{@formula} macro provide an interface that looks like \pkg{relsurv}'s in \proglang{R}.
As a reminder from \cref{tab:functions_map}, the interface on \proglang{R} for the same algorithm would be:
\begin{lstlisting}[language=R]
library(survival)
library(relsurv)
pp = rs.surv(Surv(time, status) ~ 1, 
             data = colrec, 
             ratetable = slopop, 
             method = "pohar-perme",  # by default
             add.times=1:8149)
\end{lstlisting}

By using a similar interface, we hope to help users get accustomed to the \proglang{Julia} version quicker. Note that in the \proglang{Julia} version, the mapping between the \code{RateTable} axes names and the \code{DataFrame} column names has to be perfect and cannot be corrected through the interface: you should rename your \code{DataFrame} columns before calling the fitting function. If a perfect match is not found an explicit and informative error message will be thrown.

On the other hand, the resulting excess survival curve, excess hazard function and the estimated variance of the excess hazard obtain by the Pohar Perme method on this example are given in \cref{results_pp}. 
\begin{table}[H]
  \small
  \begin{tabu} to \linewidth {>{\raggedright}X>{\raggedright}X>{\raggedright}X>{\raggedright}X>{\raggedright}X>{\raggedright}X}
\toprule
Day & $S_e$ & $\partial \Lambda_e$ & $\sigma_e$ & Lower 95\% CI & Upper 95\% CI  \\
\midrule
1 & 0.997105 &  0.00289493  & 0.000710632 &   0.995717   &   0.998495 \\
2 & 0.995214 &  0.0018967   & 0.0009186   &   0.993424   &   0.997007 \\
3 & 0.992149 &  0.00307987  & 0.00117581  &   0.989865   &   0.994438 \\
4 & 0.988748 &  0.00342788  & 0.0014077   &   0.986024   &   0.99148 \\
5 & 0.986016 &  0.00276316  & 0.0015717   &   0.982983   &   0.989058 \\
...  &  ...   &  ...         & ...         & ...          &  ...\\
8144 & 0.390212 & -0.00054201  & 0.740969    &   0.0913245  &   1.6673 \\
8145 & 0.390424 & -0.00054201  & 0.740969    &   0.091374   &   1.66821 \\
8146 & 0.390636 & -0.00054201  & 0.740969    &   0.0914235  &   1.66911 \\
8147 & 0.390847 & -0.00054201  & 0.740969    &   0.0914731  &   1.67002 \\
8148 & 0.391059 & -0.00054201  & 0.740969    &   0.0915227  &   1.67092 \\
\bottomrule
  \end{tabu}
  \caption{\label{results_pp} Output of the Pohar Perme function over a time period of 8148 days. The columns represent respectively the estimation of the excess survival, the partial excess hazard, the excess hazard's varaince and confidences intervals for the excess survival.}
\end{table}

For a better visualization of the net survival function and its confidence interval (obtained by the Delta method), the results are plotted (in \cref{pp_plot} below) using the following code: 
\begin{lstlisting}[language=Julia]
using Plots
conf_int = confint(pp; level = 0.05)
lb, ub = first.(conf_int), last.(conf_int)
plot(pp.grid, pp.Sₑ, ribbon=(pp.Sₑ - lb,ub - pp.Sₑ), xlab = "Time (days)",
     ylab = "Net survival", label = false)
\end{lstlisting}

\begin{figure}[H]
  \centering
  \includegraphics[width=\textwidth]{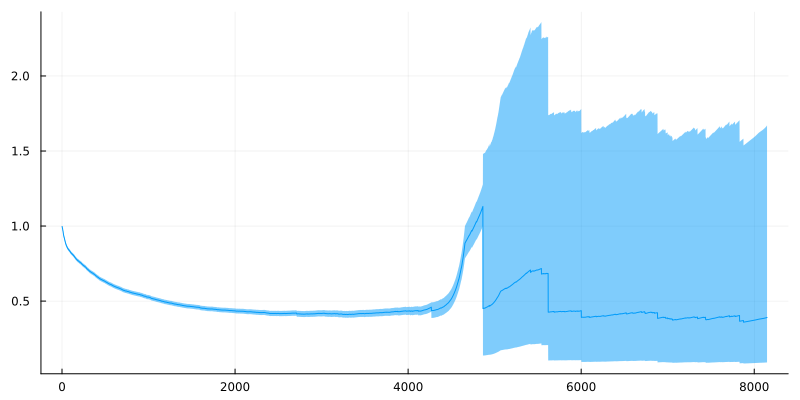}
  \caption{Pohar Perme estimator $\hat{S}_E(t)$ obtained for $t \in [1.0, 8148.0]$ (in days), with its (asymptotic) 95\% confidence interval based on $\hat{\sigma}_E$.}
  \label{pp_plot}
\end{figure}

We see on \cref{pp_plot} that the behavior of the estimator is at first very stable and accurate, until a breakpoint around day 4000 to 5000. 
After this point, not only the survival curve exhibits an unexpected sharp increase, but moreover the confidence interval significantly widens.
In fact, since $S_E(t)$ is a probability, it is supposed to stay in $[0,1]$, so an empirical confidence interval larger than that simply means that we do not have information enough to estimate \emph{anything} on later times. 

This is partially due to the age group of the patients ; given that they are elders, their population mortality rates are very high and the number of observed deaths that should be due to cancer becomes smaller and smaller.
We already mention this potential problem when introducing the net sample size in the previous section. 
To better understand the issue, we can estimate the expected net sample size with the \code{nessie} function, using a split on 10-years age groups:

\begin{lstlisting}[language=Julia]
using CategoricalArrays
breaks = [0, 45:5:90..., Inf]
colrec.agegr = cut(colrec.age./365.241, breaks)
elt, ess = nessie(@formula(Surv(time,status)~agegr), colrec, slopop)
\end{lstlisting}

This function produces two elements \code{elt} and \code{ess} that contain estimations of, respectively, the expected lifetime and expected sample size of each group w.r.t. the populational mortality only.
We present an excerpt of the excepted sample sizes computed by the function in \cref{nessie_ess}.

\begin{table}[H]
  \small
  \begin{tabu} to \linewidth {>{\raggedright}l>{\raggedright}X>{\raggedright}X>{\raggedright}X>{\raggedright}X>{\raggedright}X>{\raggedright}X>{\raggedright}X>{\raggedright}X>{\raggedright}X>{\raggedright}X>{\raggedright}X}
    \toprule
    & 0-45 & 45-50 & 50-55 & 55-60 & 60-65 & 65-70 & 70-75 & 75-80 & 80-85 & 85-90 & 90+\\
    \midrule
    1 & 246.0 & 239.0 & 397.0 & 591.0 & 879.0 & 1066.0 & 1012.0 & 725.0 & 492.0 & 261.0 & 63.0 \\
    3 & 244.81 &  236.51 & 390.68 & 576.77 & 845.07 & 1005.24 & 927.01 & 634.00 & 388.43 & 179.90 &  34.27\\
    6 & 242.64 & 232.17 & 379.87 & 552.77 & 789.94 & 906.26 & 792.11 & 493.90 & 251.10 & 91.47 & 11.44 \\
    9 & 240.07 & 227.06 & 367.45 & 526.16 & 729.33 & 799.30 & 647.10 & 341.67 & 146.22 & 37.99 & 0.00 \\
    12 & 237.08 & 220.81 & 354.20 & 497.35 & 664.52 & 681.46 & 490.74 & 201.94 & 74.90 & 11.33 & 0.00 \\
    15 & 233.58 & 214.26 & 339.38 & 465.60 & 593.48 & 559.64 & 336.49 & 119.31 & 29.79 & 2.25 & 0.00 \\
    18 & 228.20 & 204.57 & 316.61 & 416.98 & 489.08 & 395.58 & 185.61 & 45.31 & 6.35 & 0.00 & 0.00 \\
    21 & 223.29 & 195.68 & 295.69 & 372.81 & 397.36 & 271.53 & 97.70 &  16.80 & 0.00 & 0.00 & 0.00 \\
    22 & 221.40 & 192.26 & 287.52 & 355.32 & 362.99 & 230.27 & 73.84 & 0.00 & 0.00 & 0.00 & 0.00 \\
    \bottomrule
  \end{tabu}
  \caption{\label{nessie_ess} Value of $\mathbb E\left(n_E(t)\right)$ given in the \code{ess} object, as produced by the \code{nessie} function. It represents estimation of the net expected sample size per age groups (different columns for different age groups), at different future time points (in rows, to be read in years).}
\end{table}

\Cref{nessie_ess} clearly shows that the expected sample sizes of later age groups can become very low at later times.
Note that if the table shows zeros, this is a rounding artifact as the population mortality is a continuous random variable and does not have a maximum, its expectation cannot then be zero. 
Nevertheless, these numbers can be thought as rough estimates of statistical sample sizes for the estimation of net survival.
Then, this dwindling data from older participants directly affects the accuracy of estimations over time.

The crude mortality function can also help with understanding the instability of the estimator shown in \cref{pp_plot}.

Note that in \pkg{relsurv}, crude mortality has to be queried via the \code{cmp.rel} function as such:

\begin{lstlisting}[language=R]
crude_mortality = cmp.rel(Surv(time, stat) ~ 1, colrec, slopop)
\end{lstlisting}
In our implementation, the calling syntax is a bit more transparent:
\begin{lstlisting}[language=Julia]
crude_mortality = CrudeMortality(pp)
\end{lstlisting}
It produces the table given in \cref{cm_output}. 
\begin{table}[H]
  \small
  \begin{tabu} to \linewidth {>{\raggedright}X>{\raggedright}X>{\raggedright}X>{\raggedright}X}
    \toprule
$t$ & $1 - \hat{S}_O(t)$  & $\hat{M}_E(t)$   & $\hat{M}_P(t)$ \\
\midrule
1 & 0.00300587 & 0.0028862  & 0.000119666 \\
2 & 0.00501215 & 0.00477337 & 0.000238784 \\
3 & 0.00818519 & 0.00782796 & 0.000357239 \\
...   &     ...     &     ...      &      ... \\
8146 & 0.913805   & 0.394828   & 0.518977 \\
8147 & 0.913805   & 0.394821   & 0.518984 \\
8148 & 0.913805   & 0.394814   & 0.518991 \\
\bottomrule
  \end{tabu} 
  \caption{\label{cm_output} Output of the crude mortality function: Estimated crude excess mortality $\hat{M}_E(t)$ and crude population mortality $\hat{M}_P(t)$ functions for $t \in [1.0, 8148.0]$ (in days). The second column gives $1 - \hat{S}_O(t) = \hat M_E(t) + \hat M_P(t)$.}
\end{table}
The graph in \cref{fig:crude_plot} represents the results shown in \cref{cm_output} and marks the fluctuations of the crude mortality rate separated by cause as time progresses:
\begin{figure}[H]
  \centering
  \includegraphics[width=\textwidth]{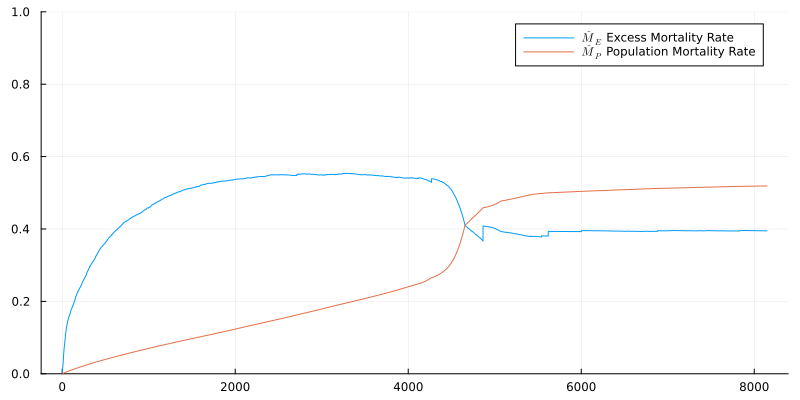}
  \caption{Estimated crude excess mortality $\hat{M}_E(t)$ and crude population mortality $\hat{M}_P(t)$ functions for $t \in [1.0, 8148.0]$ (in days).}
  \label{fig:crude_plot}
\end{figure}

As seen in \cref{fig:crude_plot}, the excess mortality rate is significantly higher at the beginning of the study than that of the population.
In other terms, the deaths recorded are more likely to be caused by colorectal cancer than other causes.
However, around the 4000 days mark (a bit more than 12 years), the excess mortality rate reaches its peak (around 55\%) before it begins to drop while the population mortality rate catches up.
By the end, the population mortality rate (around 51\%) is higher than the excess mortality rate (around 39\%).
Once again, this represents the fact that older patients are more likely to die from other causes than from the cancer.

\begin{figure}[H]
  \centering
  \includegraphics[width=\textwidth]{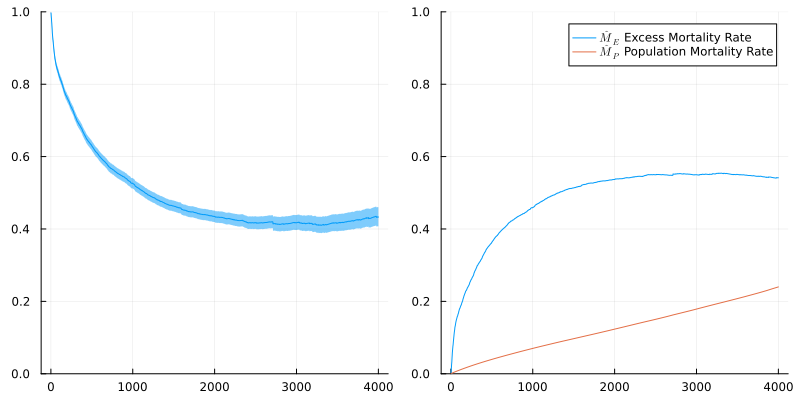}
  \caption{On the left, the Pohar Perme estimator $\hat{S}_E(t)$ for $t \in [1.0, 4000.0]$ (in days), with its (asymptotic) 95\% confidence interval. On the right, the crude mortality curves on the same timeline.}
  \label{fig:new_plot}
\end{figure}

To focus only on the first 4000 days, as indicated by these samples sizes and crude mortality analysis, we now censor patients that are still alive at that point. 
This simply corresponds to a cut on the x-axis of the net survival and crude mortality graphs, as \cref{fig:new_plot} shows.

The curve in \cref{fig:new_plot} representing the net survival function shows a more reasonable confidence interval. We still observe the unfortunate rapid decrease in the curve.

We now focus on covariates provided by our dataset and their potential impact on net survival. 
For this end, we use the Grafféo log-rank test to test the impact of different combinations of covariates.
As detailed above in \cref{npe}, this test helps compare the net survival functions between different groups.
The syntax differs a bit w.r.t. the grouping variable and the eventual stratification, as shown by these four examples: 

\begin{lstlisting}[language=Julia]
fit(GraffeoTest, @formula(Surv(time4000,status4000)~sex),               colrec, slopop)
fit(GraffeoTest, @formula(Surv(time4000,status4000)~site),              colrec, slopop)
fit(GraffeoTest, @formula(Surv(time4000,status4000)~stage),             colrec, slopop)
fit(GraffeoTest, @formula(Surv(time4000,status4000)~stage+Strata(sex)), colrec, slopop)
\end{lstlisting}

The results obtained are given in \cref{tbl:graf_output}.

\begin{table}[H]
  \small
  \begin{tabu} to \linewidth {>{\raggedright}X>{\raggedright}X>{\raggedright}X>{\raggedright}X>{\raggedright}X}
    \toprule
    Grouping variable(s) & Stratification variable(s) & Test statistic & d.o.f & p-value \\
    \midrule
    sex & - & 4.19413 & 1 & 0.0405641 \\
    site & - & 0.16802 & 1 & 0.6818780 \\
    stage & - & 949.688  & 3 & 1.47586$e^{-205}$ \\
    agegr & - & 28.7620 & 10 & 0.00136146 \\
    stage & sex & 969.781 & 7 & 4.07015$e^{-205}$ \\
    \bottomrule    
  \end{tabu}
  \caption{Output for the Grafféo log rank test function: test statistics, associated degrees of freedom and asymptotic p-value of the test are provided on the several potential grouping and stratification that are possible according to our covariates.}
  \label{tbl:graf_output} 
\end{table}

\begin{figure}[H]
  \centering
  \includegraphics[width=\textwidth]{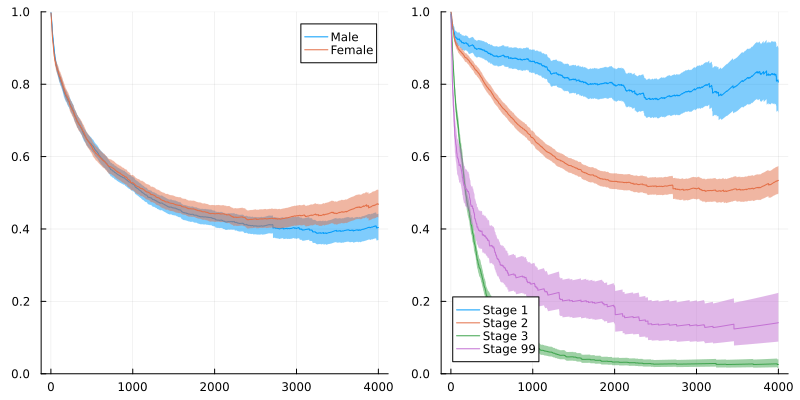}
  \caption{Pohar Perme estimator $\hat{S}_E(t)$ for $t \in [1.0, 4000.0]$ (in days), stratified by \code{sex} (on the left) and \code{stage} (on the right). Curves are represented with their respective asymptotics $95\%$ confidence intervals.}
  \label{fig:net_comparison}
\end{figure}

\Cref{tbl:graf_output} shows that at a level of $5\%$, Grafféo's log-rank test reject the null hypothesis for the \code{stage} of the cancer, the \code{sex} of the patient, but not really for the \code{site} of the cancer (colon or rectal). To get a better view of what happens in the case of the sex or the stage covariate, we can look at the net survival curves in each of the categories defined by these variables. This is represented in \cref{fig:net_comparison}.  

In \cref{fig:net_comparison}, we can see that the difference between men and women could only be due to the inaccuracy of the estimation (shown by the increasing survival curve, and the overlapping confidence intervals), that starts earlier for women than for men.
In fact, when running the Grafféo log-rank test on the whole study without censoring after 4000 days, we get that the \code{sex} variable is not significant under a 5\% error margin.
As for the stage covariate, the second graph of \cref{fig:net_comparison} shows that the more advanced the stage is the less likely it is for the patient to survive.
As a reminder, stage 99 signifies an unidentified cancer stage.
The reason its respective net survival function is drastically different from the others could be due to the small sample size of patients with unidentified cancer stages, and to the particularity of these patients: unidentified does not mean only unknown but is a special category by itself. 

\section{Conclusion} \label{conclusion}
The \pkg{relsurv} package on \proglang{R} has been the standard implementation of classic net survival estimators for about 15 years now. Our proposal, \pkg{NetSurvival.jl}, written purely in \proglang{Julia}, is much more concise, making it easier to read, maintain and validate, while being significantly faster. 

We mainly reprised classic non-parametric estimators like Ederer 1 \citep{Ederer1961}, Ederer 2 \citep{Ederer1959}, and Pohar Perme \citep{PoharPerme2012}, Grafféo's log-rank test \citep{Graffeo2016}, Cronin-Feuer estimator of crude mortality \citep{cronin2000cumulative}, alongside standard functionalities related to rate tables, such as expected lifetimes and sample sizes.
Parametric methods such as hazard regression, general hazard models, and other excess survival parametric models and regression approaches could also be easily added into the package.
The interface provided to the end user is close to \pkg{relsurv}'s, while embracing standards of the modern developing environment provided by the \proglang{Julia} language.

Albeit the numerical comparison is made using the relatively small colorectal cancer dataset \code{colrec}, the runtimes are around $20$ to $30$ times faster, depending on the method used. 
While \proglang{Julia} is a compiled language and \proglang{R} is not, we would like to stress that, since most of \pkg{relsurv} internals are in \proglang{C} (also compiled), the comparison remains fair.
The proper tests suite and the comprehensive documentation that come with the package are also a non-negligible improvement.

The most interesting part of the comparison is however in the quality of the code itself, which is as stated significantly shorter and easier to maintain.
The runtimes improvements and the shortness of the final code are mostly achieved thanks to the features of \proglang{Julia} itself (parametric types, multiple dispatch, expressiveness, etc.) and of its developing environment (convenient macros, testing interfaces, etc.): similar results can be obtained on other projects. 

\pkg{NetSurvival.jl} exists within the \proglang{JuliaSurv} organization on \proglang{GitHub}.
This organization was created in order to provide a common interface and shared internals for \proglang{Julia} packages related to survival analysis in general.
It consists of \pkg{RateTables.jl}, \pkg{SurvivalBase.jl}, \pkg{SurvivalDistributions.jl}, and \pkg{SurvivalModels.jl} as of late.
Through the construction of a common codebase for (relative) survival analysis, we hope to provide support for a wide community of researcher and engineers alike, by allowing development and optimization of previous or newly created packages.
Contributions to the organization and / or its packages are more than welcome.

\begin{minipage}{\textwidth}
\section*{Acknowledgements}

We thank the whole research team on survival analysis from SESSTIM for their kind and fruitful remarks, from the first line of code to the published version of this paper. 

\end{minipage}

\bibliography{references}

\begin{thebibliography}{49}
\newcommand{\enquote}[1]{``#1''}
\providecommand{\natexlab}[1]{#1}
\providecommand{\url}[1]{\texttt{#1}}
\providecommand{\urlprefix}{URL }
\expandafter\ifx\csname urlstyle\endcsname\relax
  \providecommand{\doi}[1]{doi:\discretionary{}{}{}#1}\else
  \providecommand{\doi}{doi:\discretionary{}{}{}\begingroup \urlstyle{rm}\Url}\fi
\providecommand{\eprint}[2][]{\url{#2}}

\bibitem[{Aalen(1978)}]{aalen1978nonparametric}
Aalen O (1978).
\newblock \enquote{Nonparametric Inference for a Family of Counting Processes.}
\newblock \emph{The Annals of Statistics}, pp. 701--726.

\bibitem[{Aalen \emph{et~al.}(2008)Aalen, Borgan, and Gjessing}]{aalen2008survival}
Aalen O, Borgan O, Gjessing H (2008).
\newblock \emph{Survival and Event History Analysis: A Process Point of View.}
\newblock Springer Science \& Business Media.

\bibitem[{Adamson and Pasta(1994)}]{Treatments1994}
Adamson GD, Pasta DJ (1994).
\newblock \enquote{Surgical Treatment of Endometriosis-Associated Infertility: Meta-Analysis Compared with Survival Analysis.}
\newblock \emph{American journal of obstetrics and gynecology}, \textbf{171}(6), 1488--1505.

\bibitem[{Allemani \emph{et~al.}(2015)Allemani, Weir, Carreira, Harewood, Spika, Wang, Bannon, Ahn, Johnson, Bonaventure \emph{et~al.}}]{allemani2015global}
Allemani C, Weir HK, Carreira H, Harewood R, Spika D, Wang XS, Bannon F, Ahn JV, Johnson CJ, Bonaventure A, \emph{et~al.} (2015).
\newblock \enquote{Global Surveillance of Cancer Survival 1995--2009: Analysis of Individual Data for 25 676 887 Patients from 279 Population-Based Registries in 67 Countries (CONCORD-2);.}
\newblock \emph{The lancet}, \textbf{385}(9972), 977--1010.

\bibitem[{Altekruse \emph{et~al.}(2010)Altekruse, Kosary, Krapcho, Neyman, Aminou, Waldron, Ruhl, Howlader, Tatalovich, Cho \emph{et~al.}}]{altekruse2010seer}
Altekruse S, Kosary C, Krapcho M, Neyman N, Aminou R, Waldron W, Ruhl J, Howlader N, Tatalovich Z, Cho H, \emph{et~al.} (2010).
\newblock \enquote{SEER Cancer Statistics Review, 1975-2007.}

\bibitem[{Andersen(2017)}]{andersen2017life}
Andersen PK (2017).
\newblock \enquote{Life Years Lost Among Patients with a Given Disease.}
\newblock \emph{Statistics in medicine}, \textbf{36}(22), 3573--3582.

\bibitem[{Andersen \emph{et~al.}(2012)Andersen, Borgan, Gill, and Keiding}]{andersen2012statistical}
Andersen PK, Borgan O, Gill RD, Keiding N (2012).
\newblock \emph{Statistical Models Based on Counting Processes.}
\newblock Springer Science \& Business Media.

\bibitem[{Besan{\c{c}}on \emph{et~al.}(2019)Besan{\c{c}}on, Papamarkou, Anthoff, Arslan, Byrne, Lin, and Pearson}]{besanccon2019distributions}
Besan{\c{c}}on M, Papamarkou T, Anthoff D, Arslan A, Byrne S, Lin D, Pearson J (2019).
\newblock \enquote{\pkg{Distributions.jl}: Definition and Modeling of Probability Distributions in the \proglang{JuliaStats} Ecosystem.}
\newblock \emph{arXiv preprint arXiv:1907.08611}.

\bibitem[{Bezanson \emph{et~al.}(2018)Bezanson, Chen, Chung, Karpinski, Shah, Vitek, and Zoubritzky}]{Bezanson2018}
Bezanson J, Chen J, Chung B, Karpinski S, Shah VB, Vitek J, Zoubritzky L (2018).
\newblock \enquote{\proglang{Julia}: Dynamism and Performance Reconciled by Design.}
\newblock \emph{Proc. ACM Program. Lang.}, \textbf{2}(OOPSLA).
\newblock \doi{10.1145/3276490}.
\newblock \urlprefix\url{https://doi.org/10.1145/3276490}.

\bibitem[{Bezanson \emph{et~al.}(2017)Bezanson, Edelman, Karpinski, and Shah}]{bezanson2017julia}
Bezanson J, Edelman A, Karpinski S, Shah VB (2017).
\newblock \enquote{\proglang{Julia}: A Fresh Approach to Numerical Computing.}
\newblock \emph{SIAM review}, \textbf{59}(1), 65--98.

\bibitem[{Bezanson \emph{et~al.}(2012)Bezanson, Karpinski, Shah, and Edelman}]{bezanson2012julia}
Bezanson J, Karpinski S, Shah VB, Edelman A (2012).
\newblock \enquote{\proglang{Julia}: A Fast Dynamic Language for Technical Computing.}
\newblock \emph{arXiv preprint arXiv:1209.5145}.

\bibitem[{Charvat and Belot(2021)}]{charvat2021mexhaz}
Charvat H, Belot A (2021).
\newblock \enquote{\pkg{Mexhaz}: An \proglang{R} Package for Fitting Flexible Hazard-Based Regression Models for Overall and Excess Mortality with a Random Effect.}
\newblock \emph{Journal of Statistical Software}, \textbf{98}, 1--36.

\bibitem[{Clerc-Urmes \emph{et~al.}(2014)Clerc-Urmes, Grzebyk, and H{\'e}delin}]{clerc2014net}
Clerc-Urmes I, Grzebyk M, H{\'e}delin G (2014).
\newblock \enquote{Net Survival Estimation with \pkg{stns}.}
\newblock \emph{The Stata Journal}, \textbf{14}(1), 87--102.

\bibitem[{Clerc-Urm{\`e}s \emph{et~al.}(2021)Clerc-Urm{\`e}s, Grzebyk, and H{\'e}delin}]{clerc2021flexrsurv}
Clerc-Urm{\`e}s I, Grzebyk M, H{\'e}delin G (2021).
\newblock \enquote{\pkg{Flexrsurv}: Flexible Relative Survival Analysis.}

\bibitem[{Collett(2023)}]{Collett2023}
Collett D (2023).
\newblock \emph{Modelling Survival Data in Medical Research.}
\newblock Chapman and Hall/CRC.

\bibitem[{Cram{\'e}r(1946)}]{cramer1946mathematical}
Cram{\'e}r H (1946).
\newblock \emph{Mathematical Methods of Statistics.}, volume~26.
\newblock Princeton university press.

\bibitem[{Cronin and Feuer(2000)}]{cronin2000cumulative}
Cronin KA, Feuer EJ (2000).
\newblock \enquote{Cumulative Cause-Specific Mortality for Cancer Patients in the Presence of Other Causes: A Crude Analogue of Relative Survival.}
\newblock \emph{Statistics in medicine}, \textbf{19}(13), 1729--1740.

\bibitem[{Danieli \emph{et~al.}(2012)Danieli, Remontet, Bossard, Roche, and Belot}]{danieli2012estimating}
Danieli C, Remontet L, Bossard N, Roche L, Belot A (2012).
\newblock \enquote{Estimating Net Survival: The Importance of Allowing for Informative Censoring.}
\newblock \emph{Statistics in medicine}, \textbf{31}(8), 775--786.

\bibitem[{Ederer(1961)}]{Ederer1961}
Ederer F (1961).
\newblock \enquote{The Relative Survival Rate: A Statistical Methodology.}
\newblock \emph{Natl. Cancer Inst. Monogr.}, \textbf{6}, 101--121.

\bibitem[{Ederer and Heise(1959)}]{Ederer1959}
Ederer F, Heise H (1959).
\newblock \enquote{The Effect of Eliminating Deaths from Cancer in General Survival Rates, Methodological Notes 11.}
\newblock \emph{End Result Evaluation Section, National Cancer Institute}.

\bibitem[{Eletti \emph{et~al.}(2022)Eletti, Marra, Quaresma, Radice, and Rubio}]{elettiUnifyingFrameworkFlexible2022}
Eletti A, Marra G, Quaresma M, Radice R, Rubio FJ (2022).
\newblock \enquote{A Unifying Framework for Flexible Excess Hazard Modelling with Applications in Cancer Epidemiology.}
\newblock \emph{Journal of the Royal Statistical Society Series C: Applied Statistics}, \textbf{71}(4), 1044--1062.
\newblock ISSN 0035-9254, 1467-9876.
\newblock \doi{10.1111/rssc.12566}.

\bibitem[{Est{\`e}ve \emph{et~al.}(1990)Est{\`e}ve, Benhamou, Croasdale, and Raymond}]{esteve1990relative}
Est{\`e}ve J, Benhamou E, Croasdale M, Raymond L (1990).
\newblock \enquote{Relative Survival and the Estimation of Net Survival: Elements for Further Discussion.}
\newblock \emph{Statistics in medicine}, \textbf{9}(5), 529--538.

\bibitem[{Fleming and Harrington(2013)}]{fleming2013counting}
Fleming TR, Harrington DP (2013).
\newblock \emph{Counting Processes and Survival Analysis.}, volume 625.
\newblock John Wiley \& Sons.

\bibitem[{Gamel and Vogel(2001)}]{gamel2001non}
Gamel JW, Vogel RL (2001).
\newblock \enquote{Non-Parametric Comparison of Relative Versus Cause-Specific Survival in Surveillance, Epidemiology and End Results (SEER) Programme Breast Cancer Patients.}
\newblock \emph{Statistical Methods in Medical Research}, \textbf{10}(5), 339--352.

\bibitem[{Goungounga \emph{et~al.}(2022)Goungounga, Mba, Graffeo, and Giorgi}]{xhaz2022}
Goungounga J, Mba D, Graffeo N, Giorgi R (2022).
\newblock \emph{\pkg{xhaz}: Excess Hazard Modelling Considering Inappropriate Mortality Rates.}
\newblock \doi{10.5281/zenodo.7010710}.
\newblock R package version 2.0.1.

\bibitem[{Grafféo \emph{et~al.}(2016)Grafféo, Castell, Belot, and Giorgi}]{Graffeo2016}
Grafféo N, Castell F, Belot A, Giorgi R (2016).
\newblock \enquote{{A Log-Rank-Type Test to Compare Net Survival Distributions.}}
\newblock \emph{Biometrics}, \textbf{72}(3), 760--769.
\newblock ISSN 0006-341X.
\newblock \doi{10.1111/biom.12477}.

\bibitem[{Gray(2014)}]{gray2014cmprsk}
Gray B (2014).
\newblock \enquote{\pkg{cmprsk}: Subdistribution Analysis of Competing Risks.}
\newblock \emph{R package version}, \textbf{2}(7).

\bibitem[{Greenwood(1926)}]{greenwood1926natural}
Greenwood M (1926).
\newblock \enquote{The Natural Duration of Cancer.}
\newblock \emph{Reports on public health and medical subjects}, \textbf{33}, 1--26.

\bibitem[{Grosclaude \emph{et~al.}(2017)Grosclaude, Roche, Fuentes-Raspall, and Larranaga}]{grosclaude2017trends}
Grosclaude P, Roche L, Fuentes-Raspall R, Larranaga N (2017).
\newblock \enquote{Trends in net survival from prostate cancer in six European Latin countries: results from the SUDCAN population-based study.}
\newblock \emph{European Journal of Cancer Prevention}, \textbf{26}, S114--S120.

\bibitem[{Guo(2010)}]{Survival2010}
Guo S (2010).
\newblock \emph{Survival Analysis.}
\newblock Oxford University Press.

\bibitem[{Hakulinen(1977)}]{Hakulinen1977}
Hakulinen T (1977).
\newblock \enquote{On Long-Term Relative Survival Rates.}
\newblock \emph{Journal of Chronic Diseases}, \textbf{30}(7), 431--443.

\bibitem[{Jackson(2016)}]{jackson2016flexsurv}
Jackson CH (2016).
\newblock \enquote{\pkg{flexsurv}: A Platform for Parametric Survival Modeling in \proglang{R}.}
\newblock \emph{Journal of statistical software}, \textbf{70}.

\bibitem[{Jenkins(2005)}]{Jenkins2005}
Jenkins SP (2005).
\newblock \enquote{Survival Analysis.}
\newblock \emph{Unpublished manuscript, Institute for Social and Economic Research, University of Essex, Colchester, UK}, \textbf{42}, 54--56.

\bibitem[{Jooste \emph{et~al.}(2013)Jooste, Grosclaude, Remontet, Launoy, Baldi, Molini{\'e}, Arveux, Bossard, Bouvier, Colonna \emph{et~al.}}]{jooste2013unbiased}
Jooste V, Grosclaude P, Remontet L, Launoy G, Baldi I, Molini{\'e} F, Arveux P, Bossard N, Bouvier AM, Colonna M, \emph{et~al.} (2013).
\newblock \enquote{Unbiased Estimates of Long-Term Net Survival of Solid Cancers in France.}
\newblock \emph{International journal of cancer}, \textbf{132}(10), 2370--2377.

\bibitem[{Kaplan and Meier(1958)}]{kaplanNonparametricEstimationIncomplete1958}
Kaplan EL, Meier P (1958).
\newblock \enquote{Nonparametric Estimation from Incomplete Observations.}
\newblock \emph{Journal of the American Statistical Association}, \textbf{53}(282), 457--481.
\newblock ISSN 0162-1459, 1537-274X.
\newblock \doi{10.1080/01621459.1958.10501452}.

\bibitem[{Karpinski \emph{et~al.}(2018)Karpinski, Carlsson, Ekre, Varela, Butterworth, and contributors}]{Pkg.jl}
Karpinski S, Carlsson K, Ekre F, Varela D, Butterworth I, contributors (2018).
\newblock \enquote{\pkg{Pkg.jl} - Package Manager for the \proglang{Julia} Programming Language.}
\newblock \url{https://github.com/JuliaLang/Pkg.jl}.

\bibitem[{Kassambara \emph{et~al.}(2017)Kassambara, Kosinski, Biecek, and Fabian}]{Kassambara2017}
Kassambara A, Kosinski M, Biecek P, Fabian S (2017).
\newblock \enquote{Package \pkg{survminer}.}
\newblock \emph{Drawing Survival Curves using ‘ggplot2’(R package version 03 1)}, \textbf{3}.

\bibitem[{Kipourou \emph{et~al.}(2022)Kipourou, Perme, Rachet, and Belot}]{kipourou2022direct}
Kipourou DK, Perme MP, Rachet B, Belot A (2022).
\newblock \enquote{Direct Modeling of the Crude Probability of Cancer Death and the Number of Life Years Lost Due to Cancer Without the Need of Cause of Death: A Pseudo-Observation Approach in the Relative Survival Setting.}
\newblock \emph{Biostatistics}, \textbf{23}(1), 101--119.

\bibitem[{Lambert and Royston(2009)}]{lambert2009further}
Lambert PC, Royston P (2009).
\newblock \enquote{Further Development of Flexible Parametric Models for Survival Analysis.}
\newblock \emph{The Stata Journal}, \textbf{9}(2), 265--290.

\bibitem[{Manevski \emph{et~al.}(2023)Manevski, Ru{\v{z}}i{\'c}~Gorenjec, Andersen, and Pohar~Perme}]{manevski2023expected}
Manevski D, Ru{\v{z}}i{\'c}~Gorenjec N, Andersen PK, Pohar~Perme M (2023).
\newblock \enquote{Expected Life Years Compared to the General Population.}
\newblock \emph{Biometrical Journal}, \textbf{65}(4), 2200070.

\bibitem[{Perkel \emph{et~al.}(2019)}]{perkel2019julia}
Perkel JM, \emph{et~al.} (2019).
\newblock \enquote{\proglang{Julia}: Come for the Syntax, Stay for the Speed.}
\newblock \emph{Nature}, \textbf{572}(7767), 141--142.

\bibitem[{Perme and Pavlic(2018)}]{PermePavlik2018}
Perme MP, Pavlic K (2018).
\newblock \enquote{Nonparametric Relative Survival Analysis with the \proglang{R} Package \pkg{relsurv}.}
\newblock \emph{Journal of Statistical Software}, \textbf{87}, 1--27.

\bibitem[{Perme \emph{et~al.}(2012)Perme, Stare, and Estève}]{PoharPerme2012}
Perme MP, Stare J, Estève J (2012).
\newblock \enquote{{On Estimation in Relative Survival.}}
\newblock \emph{Biometrics}, \textbf{68}(1), 113--120.
\newblock ISSN 0006-341X.
\newblock \doi{10.1111/j.1541-0420.2011.01640.x}.

\bibitem[{Robins \emph{et~al.}(1993)}]{robins1993information}
Robins JM, \emph{et~al.} (1993).
\newblock \enquote{Information Recovery and Bias Adjustment in Proportional Hazards Regression Analysis of Randomized Trials Using Surrogate Markers.}
\newblock In \emph{Proceedings of the Biopharmaceutical Section, American Statistical Association}, volume~24, p.~3. San Francisco CA.

\bibitem[{Sepp{\"a} \emph{et~al.}(2016)Sepp{\"a}, Hakulinen, L{\"a}{\"a}r{\"a}, and Pitk{\"a}niemi}]{seppa2016comparing}
Sepp{\"a} K, Hakulinen T, L{\"a}{\"a}r{\"a} E, Pitk{\"a}niemi J (2016).
\newblock \enquote{Comparing Net Survival Estimators of Cancer Patients.}
\newblock \emph{Statistics in Medicine}, \textbf{35}(11), 1866--1879.

\bibitem[{Spooner \emph{et~al.}(2020)Spooner, Chen, Sowmya, Sachdev, Kochan, Trollor, and Brodaty}]{spoonerComparisonMachineLearning2020}
Spooner A, Chen E, Sowmya A, Sachdev P, Kochan NA, Trollor J, Brodaty H (2020).
\newblock \enquote{A Comparison of Machine Learning Methods for Survival Analysis of High-Dimensional Clinical Data for Dementia Prediction.}
\newblock \emph{Scientific Reports}, \textbf{10}(1), 20410.
\newblock ISSN 2045-2322.
\newblock \doi{10.1038/s41598-020-77220-w}.

\bibitem[{Therneau and Lumley(2015)}]{Therneau2015}
Therneau TM, Lumley T (2015).
\newblock \enquote{Package \pkg{survival}.}
\newblock \emph{R Top Doc}, \textbf{128}(10), 28--33.

\bibitem[{Tron \emph{et~al.}(2023)Tron, Remontet, Fauvernier, Rachet, Belot, Launay, Merville, Molini{\'e}, Dejardin, Group \emph{et~al.}}]{tron2023social}
Tron L, Remontet L, Fauvernier M, Rachet B, Belot A, Launay L, Merville O, Molini{\'e} F, Dejardin O, Group F, \emph{et~al.} (2023).
\newblock \enquote{Is the Social Gradient in Net Survival Observed in France the Result of Inequalities in Cancer-Specific Mortality or Inequalities in General Mortality?}
\newblock \emph{Cancers}, \textbf{15}(3), 659.

\bibitem[{Turnbull(1976)}]{turnbull1976empirical}
Turnbull BW (1976).
\newblock \enquote{The Empirical Distribution Function with Arbitrarily Grouped, Censored and Truncated Data.}
\newblock \emph{Journal of the Royal Statistical Society: Series B (Methodological)}, \textbf{38}(3), 290--295.

\end{thebibliography}
\end{document}